\documentclass[american,aps,manuscript,aps,superscriptaddress,floatfix,final,prb,reprint,amsmath,amssymb]{revtex4-1}
\usepackage[T1]{fontenc}
\usepackage[latin9]{inputenc}
\usepackage{color}
\usepackage{babel}
\usepackage{float}
\usepackage{textcomp}
\usepackage{amsmath}
\usepackage{amssymb}
\usepackage{graphicx}
\usepackage[unicode=true,pdfusetitle,
 bookmarks=true,bookmarksnumbered=false,bookmarksopen=false,
 breaklinks=false,pdfborder={0 0 1},backref=false,colorlinks=false]
 {hyperref}

\makeatletter

\newcommand*\LyXZeroWidthSpace{\hspace{0pt}}
\providecommand{\tabularnewline}{\\}

\@ifundefined{textcolor}{}
{%
  \definecolor{BLACK}{gray}{0}
  \definecolor{WHITE}{gray}{1}
  \definecolor{RED}{rgb}{1,0,0}
  \definecolor{GREEN}{rgb}{0,1,0}
  \definecolor{BLUE}{rgb}{0,0,1}
  \definecolor{CYAN}{cmyk}{1,0,0,0}
  \definecolor{MAGENTA}{cmyk}{0,1,0,0}
  \definecolor{YELLOW}{cmyk}{0,0,1,0}
}

\usepackage{bm}

\makeatother

\begin{document}
\title{\textcolor{black}{Two-dimensional disordered Mott metal-insulator
transition}}
\author{M. Y. Suárez-Villagrán}
\affiliation{Department of Physics and Texas Center for Superconductivity, University
of Houston, Houston, Tx 77204-5005 USA}
\author{N. Mitsakos}
\affiliation{Department of Mathematics, University of Houston, Houston, Tx 77204-5008
USA}
\author{Tsung-Han Lee}
\affiliation{Department of Physics and National High Magnetic Field Laboratory,
Florida State University, Tallahassee, Florida 32306, USA}
\author{V. Dobrosavljevi\'{c}}
\affiliation{Department of Physics and National High Magnetic Field Laboratory,
Florida State University, Tallahassee, Florida 32306, USA}
\author{J. H. Miller, Jr.}
\affiliation{Department of Physics and Texas Center for Superconductivity, University
of Houston, Houston, Tx 77204-5005 USA}
\author{E. Miranda}
\affiliation{Gleb Wataghin Institute of Physics, University of Campinas (Unicamp),
Rua Sérgio Buarque de Holanda, 777, CEP 13083-859, Campinas, Brazil}
\begin{abstract}
We studied several aspects of the Mott metal-insulator transition
in the disordered case. The model on which we based our analysis is
the disordered Hubbard model, which is the simplest model capable
of capturing the Mott metal-insulator transition.\textcolor{black}{{}
We investigated this model through Statistical Dynamical Mean-Field
Theory (statDMFT). This theory is a natural extension of Dynamical
Mean-Field Theory (DMFT), which has been used with relative success
in the last several years with the purpose of describing the Mott
transition in the clean case. As is the case for the latter theory,
statDMFT incorporates the electronic correlation effects only in their
local manifestations. Disorder, on the other hand, is treated in such
a way as to incorporate Anderson localization effects. With this technique,
we analyzed the disordered two-dimensional Mott transition, using
Quantum Monte Carlo to solve the associated single-impurity problems.
We found spinodal lines at which the metal and insulator cease to
be meta-stable. We also studied spatial fluctuations of local quantities,
such as self-energy and local Green's function, and showed the appearance
of metallic regions within the insulator and vice-versa.} We carried
out an analysis of finite-size effects and showed that, in agreement
with the theorems of Imry and Ma, the first-order transition is smeared
in the thermodynamic limit. We analyzed transport properties by means
of a mapping to a random classical resistor network and calculated
both the average current and its distribution across the metal-insulator
transition.
\end{abstract}
\maketitle

\section{\textit{\emph{Introduction}}}

A phase transition at $T=0$ as a function of some external parameter
is called a quantum phase transition. It is characterized by a singular
change in the ground state of the system. Although zero temperature
is impossible to achieve, the effects of this quantum phase transition
at $T=0$ are felt at finite temperatures. Hence the importance of
studying these transitions. A quantum phase transition of great importance
is the metal-insulator transition. The distinction between metallic
and insulating behavior is only well defined at zero temperature:
while the resistivity of an insulator diverges as $T\rightarrow0$,
this transport property approaches a constant value in the case of
a metal. At finite temperatures, the resistivity is finite in both
cases. As a result, one could imagine that the metal-insulator transition
is necessary a quantum phase transition. However, several systems
exhibit an abrupt jump of resistivity, by several orders of magnitude,
at finite temperature. It is therefore natural to extend the concept
of the metal-insulator transition to the case of finite temperatures.
The metal-insulator transition has been observed in several physical
systems such as (i) doped semiconductor systems (e.g. $\text{Si:P,B}$
\citep{Rosenbaum,Paalanen}), (ii) two-dimensional electron systems
in MOSFETs (``metal-oxide-semiconductor field-effect transistors''
\citep{Anissimova}) and semiconductor heterostructures ($\textrm{GaAs/AlGaAs}$)
\citep{Hanein,Lilly}, (iii) transition metal compounds ($\textrm{V}_{2}\textrm{O}_{3},\textrm{V}\textrm{O}_{2},\textrm{NiSSe},\textrm{Nb}$)\citep{Lederer,Mazzaferro},
and (iv) organic conductors, e.g. $\kappa-(\textrm{BETD TTF})_{2}\textrm{Cu}[(\textrm{N}(\textrm{CN})_{2}]\textrm{Cl}$)\citep{Limelette}.
Many of these systems are not pure, displaying intrinsic or extrinsic
disorder.

State-of-the-art imaging techniques have enabled researchers to investigate
systems undergoing metal-insulator transitions with nanoscale resolution
\citep{Liu}. This has opened a new window into the transport properties
of disordered strongly correlated systems. Thus, it has become clear
that beneath the total resistance of a sample, the usual indicator
of the metal-insulator transition, lurks in fact an intricate inhomogeneous
landscape. Indeed, in many cases, the insulating behavior appears
as poorly conducting puddles nucleate and grow within the metallic
host and vice-versa. The first observation of this phenomenon was
made in VO$_{2}$ films on sapphire substrate by means of scattering
near-field infrared scanning spectroscopy \citep{Qazilbash,Qazilbash1}.
Stripy puddles were also observed in microcrystals of the same system
\citet{Ocallahan} as well as in films \citep{LiuPRL} with a unidirectional
substrate induced strain, revealing that the electronic degrees of
freedom are strongly coupled to the lattice ones. These studies reveal
that such non-uniform state is induced by various inhomogeneities
such as defects, strains, surfaces, cracks, etc. It is clear that
a theoretical descriptions incorporating these features in a strongly
correlated setting is called for. This is what we propose to do in
the present work.

There are some known mechanisms capable of transforming a metal into
an insulator. In the absence of interactions, a sufficiently large
level of disorder leads to the localization of the wave functions
of a particle, the so-called Anderson localization \citep{Anderson}.
A great deal is known about this mechanism. In particular, a successful
scaling theory \citep{Abrahams} has shown that all states of a particle
are localized in the presence of any level of disorder in dimensions
$d\leq2$ (considering only the case of potential scattering, ignoring
the cases of potentials with spin-orbit interaction). When $d>2$,
you must add a minimal amount of disorder for the metal to become
an insulator. This transition is known as the Anderson metal-insulator
transition. Alternatively, Mott proposed that, even in the absence
of disorder, the electron-electron interactions may in some circumstances
induce a metal-insulator transition \citep{Mott}. Although the original
Mott mechanism was essentially based on the long-term character of
the Coulomb interaction, a model with interactions of short range
proposed by Hubbard \citep{Hubbard_I,Hubbard_II,Hubbard_III} can
also exhibit a metal-insulator transition for sufficiently strong
electronic interactions when there is one electron per site of the
crystal lattice. Because of these initial proposals, this transition
induced by the interactions is known as the Mott or Mott-Hubbard transition.
The problem of understanding the conjunction of disorder and interactions
\citep{Lee,Altshuler,Castellani_Castro,Castellani_Kotliar}, despite
some progress, is still an essentially open problem.

Theoretically, several techniques have been developed to describe
the Mott transition. One of the first was made by Hubbard himself
in a series of works \citep{Hubbard_I,Hubbard_II,Hubbard_III}. His
approach consists essentially in starting with the limit in which
the electron-electron interaction is much larger than the kinetic
energy of the system (the insulator), and gradually reducing the value
of this interaction. The characteristic gap of the Mott insulator,
separating two bands of excitations called Hubbard bands, finally
closes at a critical value of the interaction $U=U_{cHubb}$ and the
system is metalized. An opposite point of view is due to Brinkman
and Rice \citep{Brinkman}. Using a variational wave function proposed
by Gutzwiller \citep{Gutzwiller,Gutzwiller1,Gutzwiller2}, they analyzed
how the correlated metal is destroyed by the increase of electronic
interactions. In this case, at a certain critical value of the interaction
$U=U_{cBR}$, the strongly correlated quasi-particles of the Fermi
liquid disappear and the system becomes an insulator. While Hubbard's
description does not adequately describe the quasi-particles of the
correlated metal, the Brinkman and Rice approach cannot correctly
predict the presence of the Hubbard bands. Both characteristics can
be observed, for example, in optical conductivity measurements, which
indicate the incompleteness of these two approaches.

The advent of the Dynamical Mean-Field Theory (DMFT) \citep{Vollhardt,georges}
enabled a description of the Mott transition that unifies the views
of Hubbard and Brinkman-Rice. DMFT is able to incorporate, for intermediate
values \LyXZeroWidthSpace \LyXZeroWidthSpace of the interaction $U$,
both the quasi-particles of the Fermi liquid at low energies and the
incoherent Hubbard bands at high energies. In this description, the
Mott transition is a first order transition, characterized by the
disappearance of the quasi-particles and leaving behind only the finite
energy excitations of the Hubbard bands. The transition is characterized
by the existence of a region of coexistence between the metallic and
the insulating phases, as in the case of supercooling and superheating
in the liquid-gas transition. Also as in the case of that transition,
the first-order phase transition line in the temperature $T$ versus
the interaction $U$ phase diagram ends at a second-order critical
point at ($T_{c}$, $U_{c}$). Below $T_{c}$, the resistivity exhibits
a jump as a function of $U$. This jump decreases with increasing
temperature and disappears at the critical point.

\textcolor{black}{The disordered Hubbard model was studied previously
with several methods: exact diagonalization \citep{PhysRevLett.86.2388},
finite- \citep{PhysRevB.55.4149,PhysRevLett.83.4610,PhysRevB.75.165113}
and zero-temperature \citep{PhysRevB.67.205112,PhysRevLett.109.026404}
quantum Monte Carlo techniques, Hartree-Fock \citep{PhysRevLett.93.126401,shinaokaimada2009,shinaokaimada2010},
variational wave functions \citep{PhysRevB.81.075106}, DMFT \citep{PhysRevB.51.10411,PhysRevLett.91.066603,PhysRevB.71.205115,PhysRevLett.102.206403}
and typical medium theory \citep{TMT,PhysRevLett.94.056404,PhysRevLett.102.146403,TMT1,PhysRevB.94.235104,PhysRevB.98.075112}.
We should mention also the related problem of the disordered Coulomb
liquid \citep{PhysRevLett.83.1826,Punnoose289}. All of these approaches,
with their strengths and weaknesses, focus on different aspects and
shed some light on this difficult problem, yet no final picture has
emerged.}

In the present work, we employ an extension of the DMFT picture of
the Mott transition that is able to incorporate non-trivial disorder
effects, the so-called Statistical Dynamical Mean Field Theory (statDMFT)
\citep{statDMFT1}. The most important features of this method are
(i) the incorporation of all Anderson localization effects (in fact,
the method is exact in the non-interacting limit), which affects the
properties of single-particle states and (ii) the incorporation of
local interaction effects, such as in the original DMFT. Non-local
interaction effects are absent in this approach. We therefore used
this method to study the effects of disorder on the Mott transition
in a two-dimensional lattice model with randomness. As in the DMFT,
a method is required for the solution of the auxiliary single-impurity
problems. We used Quantum Monte Carlo (the Hirsch-Fye algorithm \citep{Hirsch-Fye})
to solve this single-impurity problems. \textcolor{black}{Related
DMFT approaches to other types of non-homogeneous systems have been
also used in diverse contexts \citep{PhysRevB.59.2549,Miller_2001,PhysRevB.70.195342,PhysRevB.72.235108,PhysRevB.75.125114,PhysRevLett.99.046402,Snoek_2008,PhysRevLett.100.056403,PhysRevLett.101.066802,PhysRevLett.105.065301,byczuk2019}. }

Our results show that adding disorder to the system keeps the first-order
character of the transition for finite-sized systems, including the
coexistence of both metallic and insulating solutions, although the
position of the transition fluctuates spatially. The average hysteresis
loops, however, are shifted to larger values of the interaction. Furthermore,
for a given disorder realization, we observe how increasing (reducing)
the electron-electron interaction in a metallic (insulating) system
induces the the nucleation and growth of insulating (metallic) ``bubbles'',
in striking similarity to the near-field imaging results on VO$_{2}$.
As expected for a two-dimensional system, however, as the system size
increases, there is a proliferation of both metallic and insulating
``bubbles'', signaling the smearing of the first-order transition
in the thermodynamic limit. Finally, we show how we can employ a classical
random-resistor model to describe the transport on a microscopic level,
thus offering a means to analyze these highly complex inhomogeneous
states.

\section{The model}

We focus on the site-disordered Hubbard model in a two-dimensional
square lattice with first and second nearest-neighbor hopping, as
defined by the Hamiltonian:

\begin{eqnarray}
H & = & H_{0}+H_{W}+H_{U},\label{eq:hubbmodel}
\end{eqnarray}
where
\begin{eqnarray}
H_{0} & = & -\underset{\left\langle i,j\right\rangle \sigma}{\sum}t\left(c_{i\sigma}^{\dagger}c_{j\sigma}+\mathrm{h.c.}\right)\nonumber \\
 &  & -\underset{\left\langle \left\langle i,j\right\rangle \right\rangle \sigma}{\sum}\left(t^{*}c_{i\sigma}^{\dagger}c_{j\sigma}+\mathrm{h.c.}\right),\\
H_{W} & = & \underset{i\sigma}{\sum}\epsilon_{i}n_{i\sigma},
\end{eqnarray}
and

\begin{eqnarray}
H_{U} & = & U\underset{i}{\sum}\left(n_{i\uparrow}-\frac{1}{2}\right)\left(n_{i\downarrow}-\frac{1}{2}\right).
\end{eqnarray}

Here, $c_{i\sigma}^{\dagger}$ creates an electron with spin projection
$\sigma$ at site $i$ and $n_{i\sigma}=c_{i\sigma}^{\dagger}c_{i\sigma}$
is the number operator. The lattice parameter is set to $a=1$ and
a purely imaginary second nearest-neighbor hopping $t^{*}$ is introduced
in order to move the van Hove singularity away from middle of the
clean non-interacting band while at the same time maintaining particle-hole
symmetry. We will fix it to be $t^{*}=0.5it$. The particle-hole symmetric
Hubbard \textit{$U$} term accounts for a local Coulomb repulsion.
Particle-hole symmetry is destroyed only by the diagonal disorder
term $\epsilon_{j}$ which is distributed according to a uniform probability
distribution of total width $W$ centered at zero, which can be taken
as a measure of disorder strength. The clean non-interacting dispersion
relation is $\epsilon_{k}=-2t\left(\cos k_{x}+\cos k_{y}+\sin k_{x}\sin k_{y}\right)$
and the corresponding density of state is shown in Fig. 1. The half
band width is $D=\frac{\sqrt{2}+1}{2}t\approx1.207t$, which we will
take as our energy unit.

\begin{figure}[H]
\begin{centering}
\includegraphics[scale=0.25]{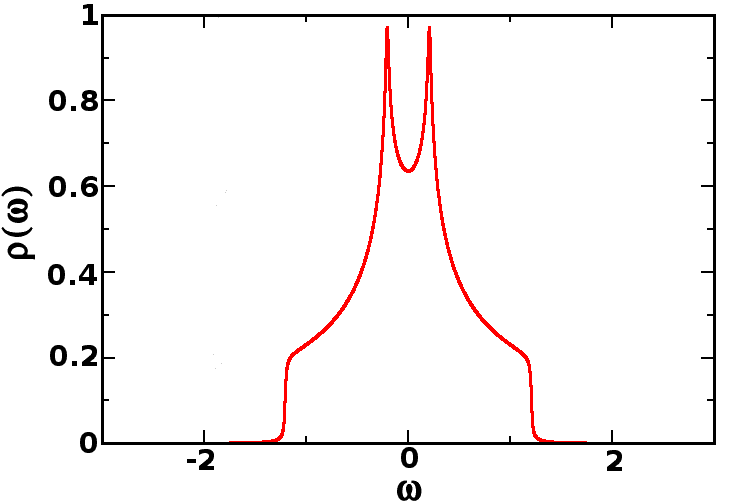}
\par\end{centering}
\begin{centering}
\caption{The clean non-interacting density of states of our model. Note that
the van Hove singularities occur away from the band center.}
\par\end{centering}
\end{figure}

\section{The statistical Dynamical Mean Field Theory (statDMFT)}

The spirit of single-site dynamical mean field theory and its descendants
is to treat exactly on-site correlations. This is achieved by assuming
a local albeit frequency-dependent self-energy. In the context of
a disordered lattice, this amounts to the following approximation
to the full self-energy 

\begin{equation}
\mathbf{\mathbf{\Sigma}}_{ij}(i\omega_{n})\rightarrow\delta_{ij}\mathbf{\mathbf{\Sigma}}_{i}(i\omega_{n}),\label{eq:self-energy}
\end{equation}
here written in its Matsubara version. Note that, although local,
the self-energy $\mathbf{\mathbf{\Sigma}}_{i}(i\omega_{n})$ varies
from site to site. The self-energy is calculated within a self-consistent
scheme as follows. Under the assumption of Eq.~(\ref{eq:self-energy}),
the local dynamics of a generic site \textit{i} is governed by the
effective action

\begin{equation}
\begin{array}{ccc}
S_{eff}^{(i)} & = & -\underset{\sigma}{\sum}\iint c_{i\sigma}^{\dagger}(\tau)g_{0}^{(i)-1}(\tau-\tau')c_{i\sigma}(\tau')d\tau d\tau'\\
 & + & U\int\left[n_{i\uparrow}(\tau)-\frac{1}{2}\right]\left[n_{i\downarrow}(\tau)-\frac{1}{2}\right]d\tau,
\end{array}\label{eq:selff}
\end{equation}
where

\begin{equation}
g_{0}^{(i)-1}(i\omega_{n})=i\omega_{n}-\epsilon_{i}-\triangle_{i}(i\omega_{n}),\label{eq:g}
\end{equation}
and $\Delta_{i}(i\omega_{n})$ is the ``cavity'' function describing
single-particle hopping to and from site $i$. The local interacting
Green's function, obtained by solving the effective action in Eq.~(\ref{eq:selff})
and defined by

\begin{equation}
G_{i}(\tau-\tau')=-\left\langle T\left[c_{i\sigma}(\tau)c_{i\sigma}^{\dagger}(\tau')\right]\right\rangle ,\label{eq:Gt}
\end{equation}
is related to the self-energy through

\begin{eqnarray}
G_{i}^{-1}(i\omega_{n}) & = & g_{0}^{(i)-1}(i\omega_{n})-\mathbf{\mathbf{\Sigma}}_{i}(i\omega_{n}).\label{eq:G}
\end{eqnarray}

\textcolor{black}{From the set of equations Eq.(\ref{eq:selff})-Eq.(\ref{eq:G})
an iterative calculational scheme can be devised. Given a finite $L\times L$
realization of the disordered lattice, we start from an initial guess
for the $L^{2}$ ``cavity'' functions $\Delta_{i}(i\omega_{n})$,
which define $L^{2}$ effective actions as given by Eq.(\ref{eq:selff})
and Eq.(\ref{eq:g}). We then use some standard impurity solver to
calculate the $L^{2}$ local interacting Green's functions from Eq.(\ref{eq:Gt})
and then find the $L^{2}$ local self-energies from Eq.(\ref{eq:G}).
This ensemble of local self-energies now has to be used to generate
updated ``cavity'' functions. This is achieved by focusing on the
single-particle lattice Green's function, which can be easily written
as a resolvent }\textcolor{black}{\emph{in the lattice site basis}}\textcolor{black}{{}
(matrices in this basis are denoted by a hat)
\begin{equation}
\widehat{G}_{lat}(i\omega_{n})=\frac{1}{i\omega_{n}\widehat{1}-\widehat{H}_{0}-\widehat{H}_{W}-\widehat{\mathbf{\Sigma}}(i\omega_{n})},\label{eq:resolvent}
\end{equation}
where $\widehat{1}$ is the unitary matrix and $\widehat{\mathbf{\Sigma}}(i\omega_{n})$
is the diagonal matrix with elements $\mathbf{\Sigma}_{i}(i\omega_{n})\delta_{i,j}$.
Physically, the renormalization introduced by interactions are encoded
in a ``shift'' of the site energies by a frequency-dependent self-energy
\begin{equation}
\epsilon_{i}\rightarrow\epsilon_{i}+\mathbf{\Sigma}_{i}(i\omega_{n}).
\end{equation}
As usual in single-site DMFT-based approaches, this renormalization
only describes local, single-particle processes. The lattice Green's
function $\widehat{G}_{lat}(i\omega_{n})$ of Eq.~\eqref{eq:resolvent}
is obtained by a frequency-by-frequency numerical inversion of the
non-Hermitian operator in the denominator. The latter can be efficiently
implemented in the site basis and the numerical inversion performed
with standard linear algebra routines. In the site basis, the diagonal
elements of $\widehat{G}_{lat}(i\omega_{n})$ are the updated local
Green's functions $G_{i}^{\left(new\right)}(i\omega_{n})$ of Eq.
(\ref{eq:G}). Therefore, the updated ``cavity'' functions can be
obtained from}

\textcolor{black}{
\begin{equation}
\Delta_{i}^{\left(new\right)}(i\omega_{n})=i\omega-\epsilon_{i}-G_{i}^{\left(new\right)-1}(i\omega_{n})-\mathbf{\mathbf{\Sigma}}_{i}(i\omega_{n}),
\end{equation}
which is then used to generate a new set of $L^{2}$ effective actions,
thus closing self-consistency loop. }The full self-consistent scheme
has been dubbed statistical dynamical mean field theory (statDMFT).
The great advantage of the method lies in its ability to track full
distributions (typically numerically) of local quantities, instead
of focusing on average, either algebraic (as in the infinite-dimensional
DFMT limit) or geometric (as in the ``typical medium theory\textcolor{black}{'')
}\citep{TMT,PhysRevLett.94.056404,PhysRevLett.102.146403,TMT1,PhysRevB.94.235104,PhysRevB.98.075112}\textcolor{black}{.
}Evidently, when interactions are turned off, the method represents
the exact diagonalization of the non-interacting disordered problem.

\textcolor{black}{It should be mentioned that originally the DMFT
of clean systems was introduced by invoking its exactness in the infinite-dimensional
limit \citet{Vollhardt}. Indeed, the infinite coordination suppresses
fluctuations in the same way as in the mean-field treatment of spin
systems. DMFT's subsequent popularization and widespread use in finite-dimensional
systems, however, has come from the realization that many strongly
correlated systems are well described within a local treatment of
correlations. In this sense, DMFT and its descendants represent the
}\textcolor{black}{\emph{optimal}}\textcolor{black}{{} implementation
of this local program. This has become especially clear in the description
of the clean Mott-Hubbard transition end-point \citet{Limelette}.
Of course, other low-temperature instabilities (like magnetism) are
especially sensitive to a finite, low dimensionality. Thus, our use
of the method in a two-dimensional case can be justified in two ways:
(a) we work close to the second-order end-point of the clean transition,
and (b) most of our focus is on the particularity of two spatial dimensions,
where the Imry-Ma effect destroys the clean first-order transition
line, as will be explained later.}

We have implemented the statDMFT approach to study the disordered
Mott transition in the two-dimensional Hubbard model at half filling.
We have focused on three different temperatures with the following
choice of parameters: $T=0.028D$ , $T=0.024D$ and $T=0.02D$ and
the value of disorder was fixed at $W=0.52D$. Since we focus on finite
temperatures, the issue of antiferromagnetic order, which only occurs
at $T=0$ in two dimensions is not important here. In our calculations,
we have used the Quantum Monte Carlo algorithm of Hirsch and Fye \citep{Hirsch-Fye}
as impurity solver. The discretization of the imaginary time axis
was set at $\triangle\tau$=0.55. At each run of the impurity solver,
the number of sweeps used to obtain the converged results was $100,000$.
The number of iterations needed to reach the full self-consistency
of the statDMFT equations was less than $50$ for well-defined metallic
or insulating solutions, but increased closer to the critical points,
where it could range from $200$ to $500$ interactions.

\section{The Mott-Hubbard phase transition and the effects of disorder}

Theoretical and experimental studies indicate that the Mott transition
belongs to the same universality class of the liquid-gas phase transition
and the Ising model\citep{Kotliar_Lange_Rozenberg}. The clean Hubbard
model Hamiltonian in the presence of a chemical potential reads

\begin{equation}
\begin{array}{ccc}
H & = & H_{0}+H_{U}-\underset{i\sigma}{\sum}\mu n_{i\sigma}\end{array}.
\end{equation}

\begin{figure}[H]
\begin{centering}
\includegraphics[scale=0.4]{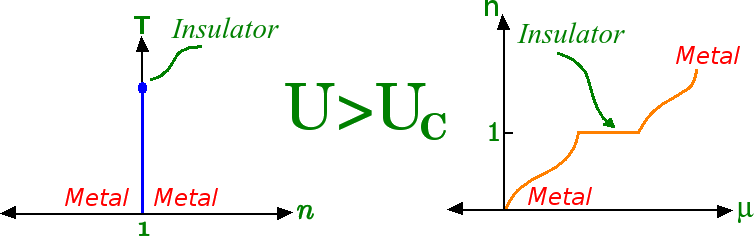}
\par\end{centering}
\caption{\label{fig:Phase-diagram}$T-n$ phase diagram for the Mott transition
and occupation number as a function of chemical potential at $T=0$
in the Hubbard model.}

\end{figure}

For values of the local interaction $U>U_{c}$, the system displays
insulating behavior when the mean occupancy number $\left\langle n\right\rangle =1$.
For values of $\left\langle n\right\rangle \neq1$, the system is
metallic. The phase diagram of the transition corresponds to a first
order transition at $n=1$, culminating at a second-order critical
point at $T_{c}$, as shown in Fig.~\ref{fig:Phase-diagram}. This
figure also shows the dependence of $n$ on the chemical potential
$\mu$ for $T=0$. Note that there is a plateau at $n=1$, since the
presence of the Mott gap makes the system incompressible $(dn/d\mu=0)$.
We would like to emphasize that this phase diagram stands in complete
analogy with the phase diagram of the Ising model in an external (longitudinal)
field $h$ when we make the correspondences $\left(n-1\right)\to m$
(where $m$ is the magnetization density) and $\mu\to h$.

Based on the above behavior, it is clear that the diagonal disorder
$H_{W}=\underset{i\sigma}{\sum}\epsilon_{i}n_{i\sigma}$ act as a
``local'' chemical potential by doping the insulator and making it
metallic in a given region for sufficiently large values of $\left|\epsilon_{i}\right|$.
By analogous reasoning to the Ising model with random fields \citep{RFIM},
if the fluctuations of $\left|\epsilon_{i}\right|$ are sufficiently
large in a certain region, the insulator is unstable with respect
to the formation of a metallic region . If $N$ is the number of sites
within the region, then the fluctuations are such that

\begin{eqnarray}
{\color{red}{\color{black}\triangle\epsilon}} & = & \sqrt{\frac{\underset{i}{\sum\epsilon_{i}^{2}}}{N-1}}\left\{ \begin{array}{cc}
{\color{red}{\color{black}\lesssim\triangle\epsilon_{c}}} & \mathtt{{\color{blue}{\color{black}Remains\:Insulating,}}}\\
{\color{red}{\color{black}\gtrsim\triangle\epsilon_{c}}} & \mathtt{{\color{blue}{\color{black}Local\:Metallization,}}}
\end{array}\right.
\end{eqnarray}
where $\triangle\epsilon_{c}\sim U$ when $T=0$. Therefore, the region
remains insulating if the size of fluctuations is less than a critical
value, above which we have local metallization. This is analogous
to the effect of the random field on the Ising model. We should mention
that a careful scaling analysis of the near-field imaging results
of \onlinecite{Qazilbash} gave strong support to a picture of the
metal-insulator transition in VO$_{2}$ based on the random field
Ising model universality class \citep{RFIMVO2,RFIM}.

There is a crucial dimension dependence to this phenomenon, however.
Indeed, the same considerations as used by Imry and Ma \citep{Imry-Ma}
in their analysis of the random field Ising model lead us to conclude
that disorder destroys the two-dimensional Metal-Insulator transition
in the thermodynamic limit. This is because below and at two dimensions
the interface energy between metal and insulator is not able to hinder
the proliferation of metallic regions in the insulator or insulating
regions in the metal. Therefore, the system breaks into various metal
and insulating regions and the phase transition is smeared. The first-order
phase transition line on the left-hand side of Fig.~\ref{fig:Phase-diagram}
is destroyed in this two-dimensional case. This is the generalization
of the Imry and Ma theorem \citep{Imry-Ma} to the Mott transition
case. 

\textcolor{black}{Finally, we need to explain how we determine whether
a certain region belongs to an insulating or a metallic ``bubble''.
The local density of states (DOS) might be a good indicator. In a
clean system, it can be obtained through the value of the local Green's
function at a particular value of the imaginary time \citep{PhysRevLett.75.312}
\begin{equation}
G_{i}\left[\tau=\frac{1}{2k_{B}T}\right].
\end{equation}
This approach, however, assumes that the local DOS does not vary appreciably
with the frequency on the scale of the temperature. This is a reasonable
assumption in a clean Hubbard model, in which the only energy scales
are $U$ and $D$. In a disordered Hubbard model, however, the local
on-site energies fluctuate in the range $\left[-W/2,W/2\right]$,
thus generating a continuum of small energy scales over which the
local DOS varies and invalidating this procedure. Another option would
be the local self-energy at low }\textcolor{black}{\emph{real}}\textcolor{black}{{}
frequencies, since 
\begin{equation}
\mathrm{Re}\Sigma_{i}\left(\omega\right)\to\begin{cases}
0 & \mathrm{metal},\\
\infty & \mathrm{insulator}.
\end{cases}\label{eq:realselfen}
\end{equation}
This indicator would require the analytical continuation from Matsubara
to real frequencies, a notably difficult task. Since this must be
performed at every lattice site, we tried to create an automated algorithm
to do this, using the usual maximum entropy and Padé techniques. However,
this proved to be very unreliable. In the end, we opted for the value
of the imaginary part of the local self-energy at the first Matsubara
frequency $\mathrm{Im}\Sigma_{i}\left(i\omega_{1}\right)$, since
it reflects the same tendency of Eq.~\eqref{eq:realselfen}, being
large in the insulator and small in the metal
\begin{equation}
\mathrm{Im}\Sigma(i\omega_{1})\sim\left\{ \begin{array}{cc}
i\omega_{1}\to0\ \mathrm{as}\ T\to0 & \mathrm{metal},\\
1/i\omega_{1}\to0\ \mathrm{as}\ T\to0 & \mathrm{insulator}.
\end{array}\right.
\end{equation}
}

\section{Transport properties}

\label{sec:Transport-properties}

\textcolor{black}{It would be useful to use the data from $\mathrm{Im}\Sigma_{i}(i\omega_{1})$
as a means to access the transport properties within statDMFT. This
is possible at $T=0$ in the non-interacting case by means of the
Landauer formalism \citep{Landauer}, through the calculation of the
transmission matrix between the edges of the system. This formalism
was later extended to interacting systems and finite temperatures
\citet{PhysRevLett.68.2512} enabling a full statDMFT calculation
of transport properties. }Nonetheless, when the transport occurs \emph{without
quantum coherence at any length scale} due to the strong inelastic
scattering, it is possible to make a classical description of the
resistivity. We will show that at the temperatures and interactions
in which we work, transport is completely incoherent and we will thus
use a network of classical resistors to calculate the relative resistance
values of the system. To our knowledge, this is the first attempt
to calculate transport within statDMFT, albeit in this incoherent
regime.

From many-body theory, there is a relation between the value of self-energy
at zero (real) frequency and wave vector on the Fermi surface and
the inelastic half-life of the particle \citep{Walecka}
\begin{equation}
\mathrm{Im}\Sigma(\left|\vec{k}\right|\sim k_{F},\omega\simeq0)\sim\frac{1}{\tau_{in}\left(\vec{k}\right)}.
\end{equation}
If $\tau_{in}(\vec{k})$ is approximately isotropic $\tau_{in}(\vec{k})\rightarrow\tau_{in}$,
the Kubo formula gives us the conductivity and, therefore, the resistivity
as

\begin{equation}
\rho\sim\frac{1}{\tau_{in}}\propto\mathrm{Im}\Sigma(\left|\vec{k}\right|\sim k_{F},\omega\simeq0).\label{eq:resistivity}
\end{equation}
In fact, from the Drude formula,

\begin{equation}
\rho=\frac{m}{ne\text{\texttwosuperior}\tau},
\end{equation}
in accordance with Eq.~\eqref{eq:resistivity} if the transport is
completely dominated by inelastic processes. In this case, we define
the free inelastic mean free path as being

\begin{equation}
l_{in}=v_{F}\tau_{in},\label{eq:caminho_livre}
\end{equation}
where $v_{F}$ is the Fermi velocity. For $l\gtrsim l_{in}$ the transport
is incoherent because inelastic scattering destroys the ``memory''
of the quantum phase of the electronic wave function. At these scales,
we can describe the transport classically. We can estimate the Fermi
velocity by $v_{F}\sim\frac{E_{F}}{k_{F}}$ where $E_{F}$ is the
Fermi energy. In the Hubbard model, $E_{F}$ can be taken as the half-bandwidth
$D$ in the case of half-filling. Finally, using $k_{F}\sim1/a$ where
$a$ is the lattice parameter we have $v_{F}\sim aD$. Therefore,
from Eqs. (\ref{eq:resistivity}) and (\ref{eq:caminho_livre}) we
obtain

\begin{eqnarray}
l_{in} & = & v_{F}\tau_{in}=\frac{aD}{\mathrm{Im}\Sigma}\Rightarrow\frac{l_{in}}{a}=\frac{D}{\mathrm{Im}\Sigma}.\label{eq:lin}
\end{eqnarray}
As we will show later, for the temperatures we are focusing on here,
$l_{in}\lesssim a$ and the transport is completely incoherent thus
allowing for a classical description.

\begin{figure}[H]
\centering{}\includegraphics[scale=0.4]{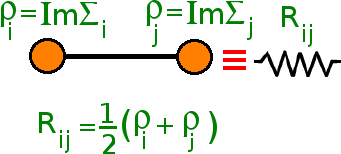}\caption{\label{fig:Resistor}Square resistor network in which each resistor
couples two neighboring sites.}
\end{figure}

Supposing now we are in the regime where $l_{in}\sim a$, let us now
describe how we can replace the interacting electron system by a network
of classical resistors. First, each site in the original network gets
associated with a local resistivity value $\rho_{i}=\mathrm{Im}\Sigma_{i}(\omega_{1})\sim1/\tau_{in}\left(i\right)$.
The bond between two nearest neighbors $i$ and $j$ is then replaced
by a resistor whose value is the average value of the resistivities
of the two sites $\rho_{i}$ and $\rho_{j}$
\begin{equation}
R_{ij}=\frac{1}{2}(\rho_{i}+\rho_{j}),\label{eq:R}
\end{equation}
 as shown in Fig. \ref{fig:Resistor}.

\begin{figure}[H]
\centering{}\includegraphics[scale=0.38]{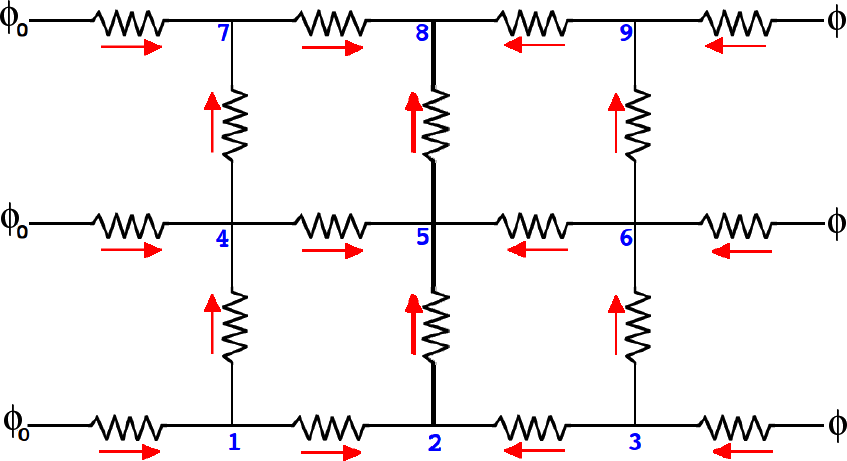}\caption{\label{fig:Resistor-1}Resistors associated to a 3 x 3 square network.}
\end{figure}

The various resistors are connected through the geometry of the network.
At the ends of the network external resistors are placed that are
connected to external voltages $\phi_{i}$. These external resistor
values are given by the resistivities at the edge sites. The external
voltages are fixed as $\phi_{0}$ at the left edge and $\phi$ at
the right edge. The network of resistors has the form shown in Fig.
\ref{fig:Resistor-1} for the particular case of a $3x3$ network.
The internal voltages and currents, which pass through each resistor
are unknown and need to be determined using electrical circuit theory.

In general, for a network with $LxL$ sites, the total number of nodes
is $2L+L{{}^2}$, with $L^{2}$ internal nodes and $2L$ external
nodes. The total number of resistors is $2L{{}^2}$. We need to find
the $2L^{2}$ currents $I_{ij}$ that cross the resistors and the
$L^{2}$ voltages at each internal node. In all, therefore, there
are $3L{{}^2}$ unknowns. The current at the inner nodes is conserved
(Kirchhoff's law) providing $L^{2}$ equations

\begin{equation}
\sum_{j}I_{ij}=0.
\end{equation}
For each resistor, we apply Ohm's law

\begin{equation}
I_{ij}=\frac{V_{i}-V_{j}}{R_{ij}},\label{eq:ohm}
\end{equation}
which gives us $2L^{2}$ equations. We therefore have a total of $3L^{2}$
equations for $3L^{2}$ unknowns. We found the solutions numerically.

\section{Results and Discussion}

To describe the Mott transition in two dimensions we use the Hubbard
Hamiltonian in the two dimensional square lattice given in Eq.~\eqref{eq:hubbmodel}.
Several studies of the clean case have established the first-order
nature of the transition at finite temperatures below $T_{c}$, with
the corresponding coexistence region and associated hysteresis \citep{vucicevic,terletska,vlad_book,miranda}.
For a finite-size system, we expect the hysteresis to survive. Therefore,
we have to allow for the convergence of both stable and meta-stable
solutions to the statDMFT equations. We thus start from initial $U$
values which are safely outside the coexistence region, either in
metallic or in the insulating phase. For a given interaction value
$U_{0}$ the values of $G_{i}(i\omega_{n})$ are found once convergence
has been achieved. The results of $G_{i}(i\omega_{n})$ for $U_{0}$
are used as an initial guess for $G_{i}(i\omega_{n})$ at $U_{1}=U_{0}+\Delta U$
(going from the metal to the insulator) or $U_{1}=U_{0}-\Delta U$
(going from the insulator to the metal). The new results for $G_{i}(i\omega_{n})$
are used to generate the $G_{i}(i\omega_{n})$ corresponding to the
next value $U_{2}$ and thus consecutively, doing a scan of $U$ values
from the metal to the insulator or from the insulator to the metal.

\begin{figure}[H]
\begin{centering}
\includegraphics[scale=0.35]{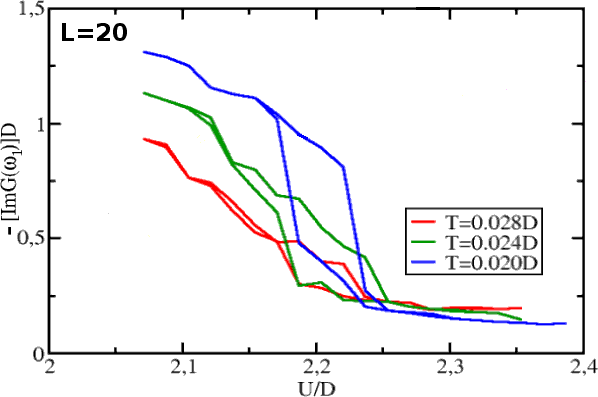}
\par\end{centering}
\centering{}\caption{\label{fig:clean_case}Hysteresis curves for different temperature
values, below the critical point of clean Mott transition. As the
temperature increases, the hysteresis loops become smaller.}
\end{figure}

As $U$ is scanned an abrupt jump is observed in $ImG(\omega_{1})$.
We define $U_{c1}$ to be this critical jump value when going from
insulator to metal and $U_{c2}$ to represent the value when going
from metal to insulator. These are the so-called spinodals. The difference
of paths traveled going from metal to insulator and from insulator
to metal defines hysteresis curves, such as shown in Fig \ref{fig:clean_case},
for the clean case. The region that is contained between $U_{c1}$
and $U_{c2}$ is the coexistence region, in which one of the solutions
is only meta-stable. In the coexistence region, for a value of $U$,
it is possible to find the two behaviors, metallic and insulating.

\begin{figure}[H]
\begin{centering}
\includegraphics[scale=0.17]{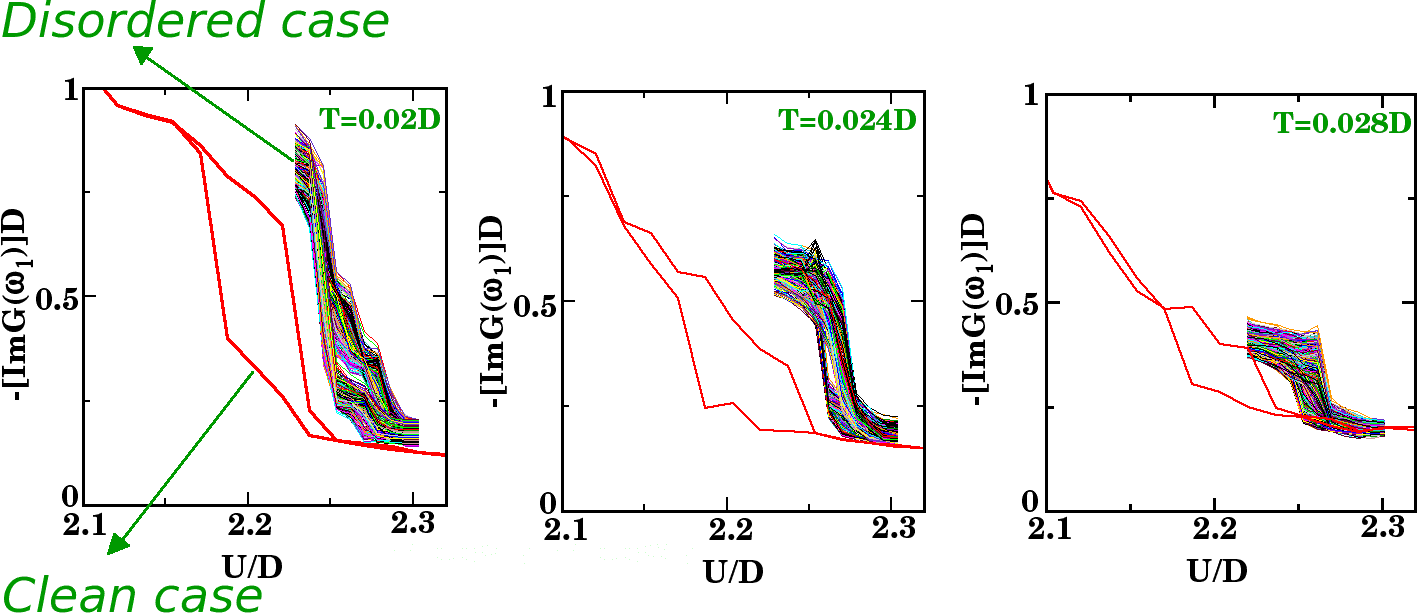}
\par\end{centering}
\centering{}\caption{\label{fig:disordered_case}Hysteresis curves for different values
of temperature. Notice how the disorder shifts the hysteresis curve
to higher values of interaction energy, while the coexistence region
shrinks.}
\end{figure}

The disordered case is shown in the Fig. \ref{fig:disordered_case}.
Note that, since $G_{i}(i\omega_{n})$ now fluctuates spatially we
have shown all the $L^{2}$ curves for $-\mathrm{Im}G_{i}\left(i\omega_{1}\right)$.
Although there are many curves, a clear hysteretic behavior is apparent,
especially at the lowest temperatures. Besides, adding disorder causes
a shift in the hysteresis curves to higher interaction values and
the coexistence region shrinks in size.

\begin{figure}[H]
\begin{centering}
\includegraphics[scale=0.4]{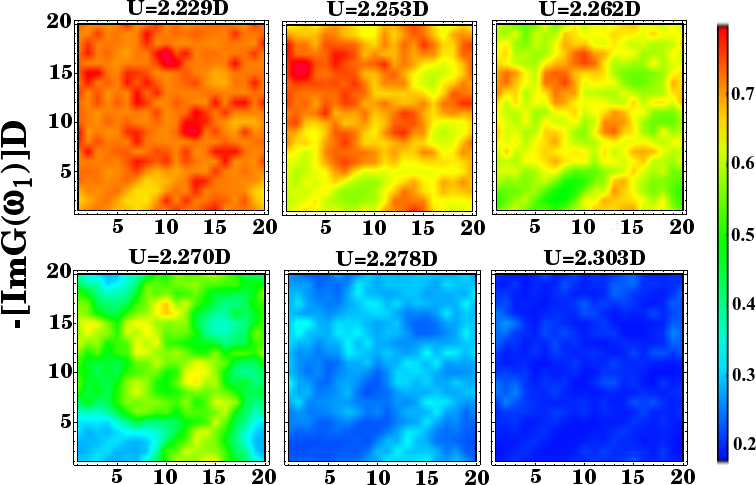}
\par\end{centering}
\centering{}\caption{\label{fig:metal_insulator}Imaginary part of the Green's function
at the first Matsubara frequency for each site of the square lattice
with $T=0.024D$ in the neighborhood of the Mott transition when going
from the metal to the insulator. See video mi.avi at Supplementary
Material \citep{SMaterial}.}
\end{figure}

\begin{figure}[H]
\begin{centering}
\includegraphics[scale=0.4]{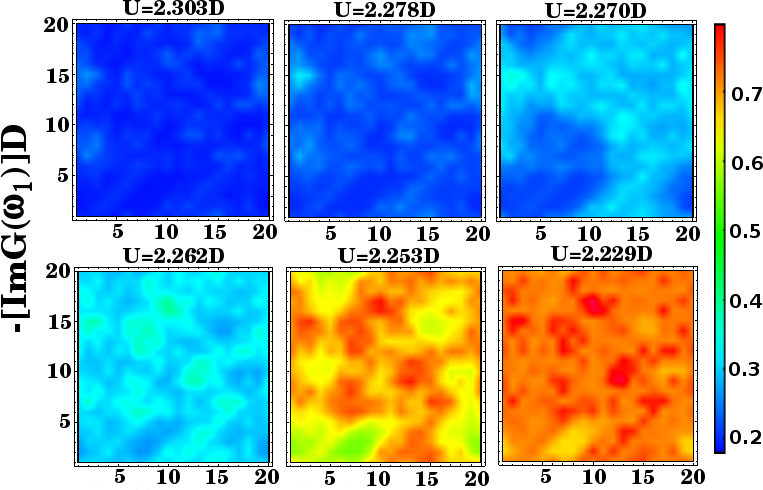}
\par\end{centering}
\centering{}\caption{\label{fig:Insulator_metal}Imaginary part of the Green's function
at the first Matsubara frequency for each site of the lattice for
$T=0.024D$ and a scan of values \LyXZeroWidthSpace \LyXZeroWidthSpace of
$U$ in the neighborhood of the Mott transition going from the insulator
to the metal. See video im.avi at Supplementary Material \citep{SMaterial}.}
\end{figure}

Let us now focus on the vicinity of the Mott transition. For each
value of $U$ in a scan of values \LyXZeroWidthSpace \LyXZeroWidthSpace where
$\Delta U=0.008D$, a map is obtained representing the spatial behavior
of the imaginary part of the Green's function at the first Matsubara
frequency. Fig \ref{fig:metal_insulator} shows these results for
$T=0.024D$ going from the metal to the insulator (\textcolor{black}{Results
for other values of temperature, can be found in the Supplementary
Material} \citep{SMaterial}). The color scale is organized so that
the largest value of $-\mathrm{Im}G(\omega_{1})$ corresponds to red
and the smallest values to blue. As the value of the local interaction
changes, the spatial configuration in the lattice changes. The system,
which initially was a metal with significant spatial homogeneity,
begins to show ``bubbles'' corresponding to insulating regions. Finally,
these coalesce to form a rather homogeneous insulator. Similarly,
starting with high local electron-electron interaction it is observed
that, as the value of the local interactions decrease, metallic \textquotedbl bubbles\textquotedbl{}
appear until the lattice becomes metallic, as shown in Fig \ref{fig:Insulator_metal}.
Comparing the same intermediate values of $U$ in the two figures,
we can easily distinguish the two coexisting solutions.

\subsection{Critical behavior of the Mott transition}

\begin{figure}[H]
\begin{centering}
\includegraphics[scale=0.28]{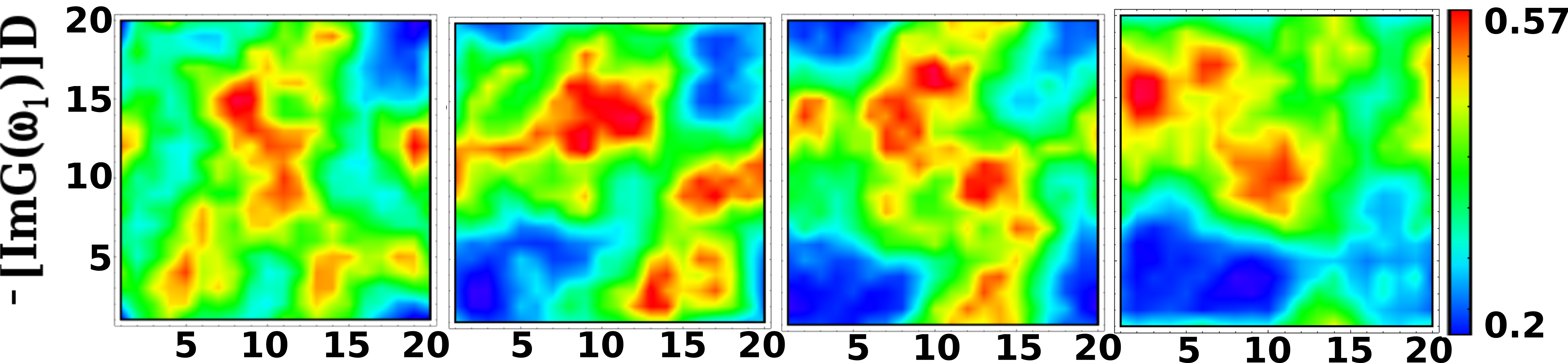}
\par\end{centering}
\caption{\label{random_seed}For four different realizations of disorder, we
show $\mathrm{Im}G_{i}(\omega_{1})$ via a color scale. The red color
represents the lattice sites that have metallic behavior. Insulating
behavior corresponds to blue regions. We used $W=0.52D$, $U=2.27D$
and $T=0.024D$.}

\end{figure}

We now focus on the correlations between the fluctuations of the bare
disorder and the local order parameter of the Mott transition. Fig.
\ref{random_seed} shows the spatial patterns of the local order parameter
$\mathrm{-Im}G_{i}(i\omega_{1})$ for four different disorder realizations
at $W=0.52D$, $U=2.27D$ and $T=0.024D$. The range of variations
for each disorder realization is between $0.20<[-\mathrm{Im}G(\omega_{1})]D<0.57$.
The red color represents the regions with greater metallic behavior
and blue regions represent the insulator. For convenience, let us
define an essentially metallic region as one in which the condition
$[-ImG(\omega_{1})]D>0.45$ is satisfied. Analogously, essentially
insulating regions are defined as those in which $[-ImG(\omega_{1})]D<0.27$.
In each of these regions we calculate the relative local fluctuation
of the disorder $\Delta\epsilon/\Delta\epsilon_{distr}$, where we
take $\bar{\epsilon}=0$, $\Delta\epsilon=\sqrt{\frac{\overset{N}{\underset{i=1}{\sum}}\epsilon_{i}^{2}}{N-1}}$
and the standard deviation of the bare distribution is $\Delta\epsilon_{distr}=\sqrt{\frac{W^{2}}{12}}=0.15D$.

Table \ref{tab:Fluctuations} shows the results for the $\Delta\epsilon/\Delta\epsilon_{distr}$
for each one of the regions. In the insulating regions, where $0.20\le[-ImG(\omega_{1})]D\le0.27$,
the values of $\Delta\epsilon/\Delta\epsilon_{distr}$ are always
smaller than 1. The values $\Delta\epsilon/\Delta\epsilon_{distr}$
in the metallic regions, where $0.45<[-ImG(\omega_{1})]D<0.57$, on
the other hand are all larger than 1. There is a strong correlation
between small (large) values of $\Delta\epsilon$ and insulating behavior
(metallic). In the regions with intermediate behavior $\Delta\epsilon\simeq\Delta\epsilon_{distr}$.
These results are in qualitative agreement with the analysis of reference
\onlinecite{RFIMVO2} of the insulating and metallic puddles of VO$_{2}$,
which showed that the observed scaling behavior is best described
by the critical random field Ising model. Unfortunately, we cannot
access very large lattice sizes in order to be able to do a full scaling
analysis of the puddle sizes.

\begin{table}[H]
\begin{centering}
\textcolor{black}{\scriptsize{}}%
\begin{tabular}{|c|c|c|c|}
\hline 
 & \textcolor{black}{\scriptsize{}Insulator} & \textcolor{black}{\scriptsize{}Intermediate regime} & \textcolor{black}{\scriptsize{}Metal}\tabularnewline
\hline 
\hline 
\textcolor{black}{\scriptsize{}Intervals} & \textcolor{black}{\tiny{}$A$} & \textcolor{black}{\tiny{}$B$} & \textcolor{black}{\tiny{}$C$}\tabularnewline
\hline 
\textcolor{black}{\scriptsize{}Number 1$\Delta\epsilon/\Delta\epsilon_{distr}$} & \textcolor{black}{\scriptsize{}$0.79$} & \textcolor{black}{\scriptsize{}$0.88$} & \textcolor{black}{\scriptsize{}$1.28$}\tabularnewline
\hline 
\textcolor{black}{\scriptsize{}Number of sites} & \textcolor{black}{\scriptsize{}$7$} & \textcolor{black}{\scriptsize{}$273$} & \textcolor{black}{\scriptsize{}$120$}\tabularnewline
\hline 
\textcolor{black}{\scriptsize{}Number 2 $\Delta\epsilon/\Delta\epsilon_{distr}$} & \textcolor{black}{\scriptsize{}$0.75$} & \textcolor{black}{\scriptsize{}$0.93$} & \textcolor{black}{\scriptsize{}$1.36$}\tabularnewline
\hline 
\textcolor{black}{\scriptsize{}Number of sites} & \textcolor{black}{\scriptsize{}$75$} & \textcolor{black}{\scriptsize{}$262$} & \textcolor{black}{\scriptsize{}$63$}\tabularnewline
\hline 
\textcolor{black}{\scriptsize{}Number 3 $\Delta\epsilon/\Delta\epsilon_{distr}$} & \textcolor{black}{\scriptsize{}$0.81$} & \textcolor{black}{\scriptsize{}$0.90$} & \textcolor{black}{\scriptsize{}$1.33$}\tabularnewline
\hline 
\textcolor{black}{\scriptsize{}Number of sites} & \textcolor{black}{\scriptsize{}$74$} & \textcolor{black}{\scriptsize{}$264$} & \textcolor{black}{\scriptsize{}$62$}\tabularnewline
\hline 
\textcolor{black}{\scriptsize{}Number 4 $\Delta\epsilon/\Delta\epsilon_{distr}$} & \textcolor{black}{\scriptsize{}$0.73$} & \textcolor{black}{\scriptsize{}$0.93$} & \textcolor{black}{\scriptsize{}$1.30$}\tabularnewline
\hline 
\textcolor{black}{\scriptsize{}Number of sites} & \textcolor{black}{\scriptsize{}$76$} & \textcolor{black}{\scriptsize{}$264$} & \textcolor{black}{\scriptsize{}$60$}\tabularnewline
\hline 
\end{tabular}\caption{\label{tab:Fluctuations}Fluctuations of $\Delta\epsilon$ for the
insulating, metallic and intermediate regions, corresponding to the
different disorder realizations of Fig. \ref{random_seed}. Here $A=0.20\protect\leq[-\mathrm{Im}G(\omega_{1})]D\protect\leq0.27$,
$B=0.27<[-\mathrm{Im}G(\omega_{1})]D<0.45$ and $C=0.45\protect\leq[-\mathrm{Im}G(\omega_{1})]D\protect\leq0.57$}
\par\end{centering}
\end{table}

\subsection{Finite-size effects}

The Mott transition is smeared in the thermodynamic limit in $d=2$,
since there is a proliferation of metallic and insulating regions
when $L\rightarrow\infty$. Let us now study how our results change
as we increase $L$.

\begin{figure}[H]
\begin{centering}
\includegraphics[scale=0.18]{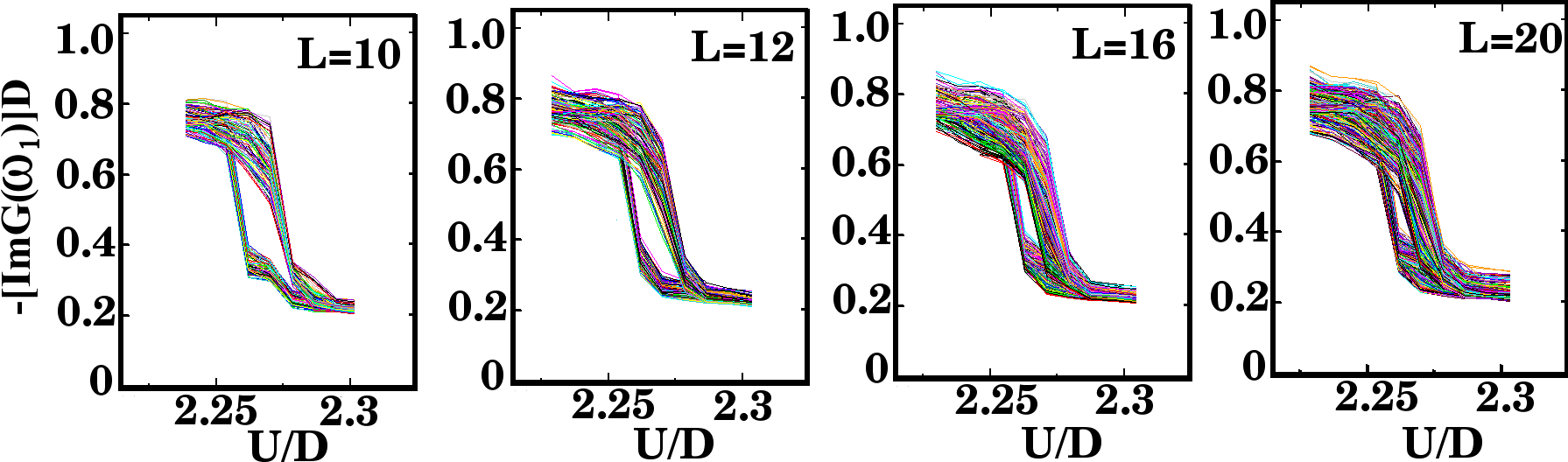}
\par\end{centering}
\centering{}\caption{\label{fig:Hysteresis-loops}Hysteresis loops for different lattice
sizes, below the critical point of Mott transition. As the temperature
increases, the hysteresis loops become smaller.}
\end{figure}

Fig \ref{fig:Hysteresis-loops} shows a set of hysteresis curves for
different lattices sizes, at $T=0.024D$ and $W=0.52D$. It is observed
that the $U/D$ values \LyXZeroWidthSpace \LyXZeroWidthSpace at which
the Mott transition occurs are the same independently of the lattice
size, while the values \LyXZeroWidthSpace \LyXZeroWidthSpace of $[-\mathrm{Im}G(\omega_{1})]D$
remain between $0.2$ and $0.8$. Note that in the case where we have
a square lattice of $10\times10$ sites, the size of the coexistence
region is larger than for larger lattice sizes. Furthermore, the first-order
Mott transition becomes a ``rounded'' transition as $L\rightarrow\infty$,
in accordance with the generalized Imry and Ma theorem for the disordered
Hubbard model. Unfortunately, it is computationally very difficult
to obtain results for $L>20$.

\begin{figure}[H]
\begin{centering}
\includegraphics[scale=0.17]{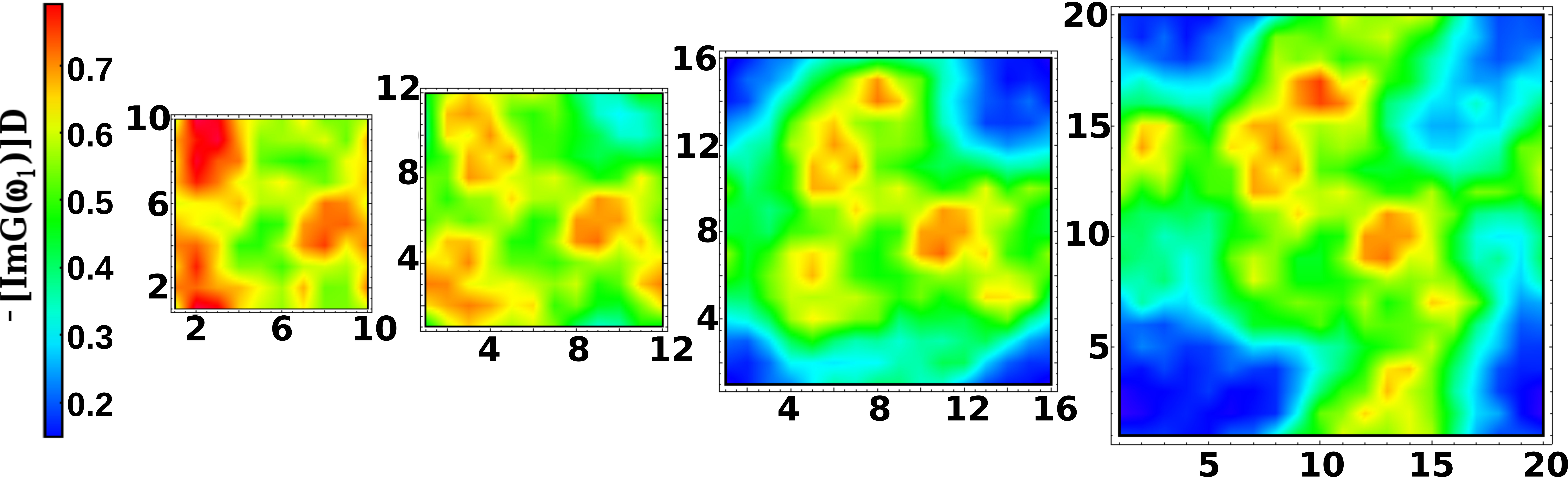}
\par\end{centering}
\centering{}\caption{\label{fig:nucleation_size_dependence}Finite-size effects: Spatial
pattern of $\mathrm{Im}G_{i}\left(i\omega_{1}\right)$ for different
sizes of the square lattice. Metallic and insulating bubbles proliferate
as the lattice increases at $U=2.27D$ in the upper branch of the
hysteresis loop. }
\end{figure}

Now we focus on a particular Coulomb interaction value $U=2.27D$
and we analyze the spatial pattern of $\mathrm{Im}G_{i}\left(i\omega_{1}\right)$,
as shown in Fig.\ref{fig:nucleation_size_dependence} for $T=0.024D$,
in the upper branch of the hysteresis loop (\textcolor{black}{for
other temperature values see the Supplementary Material} \citep{SMaterial}).
Notice how as the size of the lattice increases, metallic regions
persist at the same positions, but also note the appearance of insulating
bubbles. In the thermodynamic limit we have the proliferation of metallic
and insulating regions and the complete smearing of the transition.
Again, we note that both the inhomogeneous state with coexisting bubbles
and the accordance with the Imry-Ma theorem are in agreement with
the picture of the transition in VO$_{2}$ as being in the same universality
class as the random field Ising model \citep{RFIMVO2}.

\textcolor{black}{At this point, let us make some remarks regarding
the difference between the clean and disordered cases. In the clean
Hubbard model close to the Mott transition, thermal fluctuations also
generate metallic and insulating bubbles \citep{PhysRevLett.100.026408,HelmesPhDThesis,LiuPhDThesis}.
Their frequency and size are determined by a Boltzmann factor. In
the disordered case, however, the bubbles are nucleated by the interplay
of both temperature and local fluctuations of the disorder potential,
with the latter playing a dominant role. That can be roughly gleaned
from the persistence of the bubble landscape as the temperature is
varied with a fixed disorder realization (see Fig. 1 of the Supplemental
Material \citep{SMaterial}). Furthermore, the correlation between
the size of the site-energy fluctuations and the nature of the bubbles
(Table I) corroborates this conclusion. Finally, in the 2D case we
focus on here, the first-order transition is destroyed by disorder.
We conclude that the nature and features of the bubbles are very different
in the clean and the disordered cases.}

\subsection{Transport in the lattice}

\begin{figure}[H]
\begin{centering}
\includegraphics[scale=0.4]{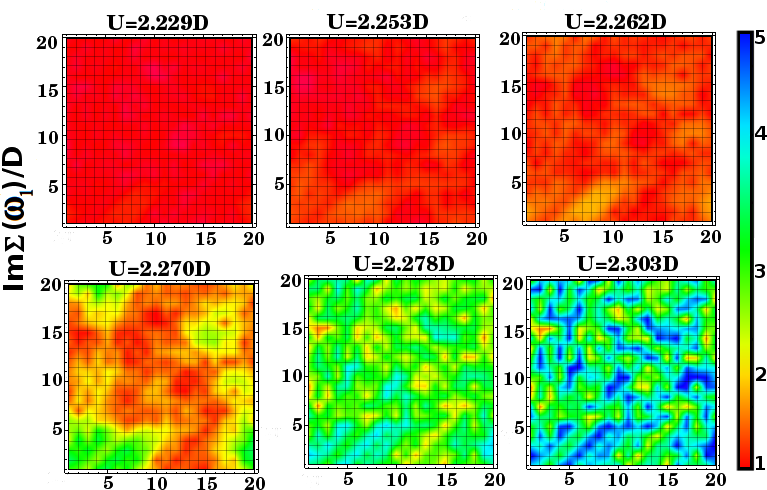}
\par\end{centering}
\caption{\label{fig:self-energy}Imaginary part of self-energy for the first
Matsubara frequency for $T=0.024D$.}
\end{figure}

To study the transport properties we analyzed the quantity $l_{in}=\frac{D}{\mathrm{Im}\Sigma}$
as described in Eq. (\ref{eq:lin}) through statDMFT. The lowest frequency
that can be used in this case is the first Matsubara frequency. Thus,
we use $\Sigma_{i}(i\omega_{1})$ as an estimate of $\Sigma_{i}(\omega\rightarrow0)$.
Fig. \ref{fig:self-energy}, depicts the value $\frac{Im\Sigma(i\omega_{1})}{D}$,
for $T=0.024D$ and $W=0.52D$, for each site of the lattice and value
of interaction $U$, in the vicinity of the Mott transition. Notice
that $0.98\leq\frac{Im\Sigma(i\omega_{1})}{D}\leq5.09$, or $0.2\lesssim\frac{D}{Im\Sigma(i\omega_{1})}\lesssim1$.
According to Eq.~\eqref{eq:lin} from Section \ref{sec:Transport-properties},
this corresponds to $l_{in}\lesssim a$. Therefore, we can describe
transport classically at all scales.

\begin{figure}[H]
\begin{centering}
\includegraphics[scale=0.4]{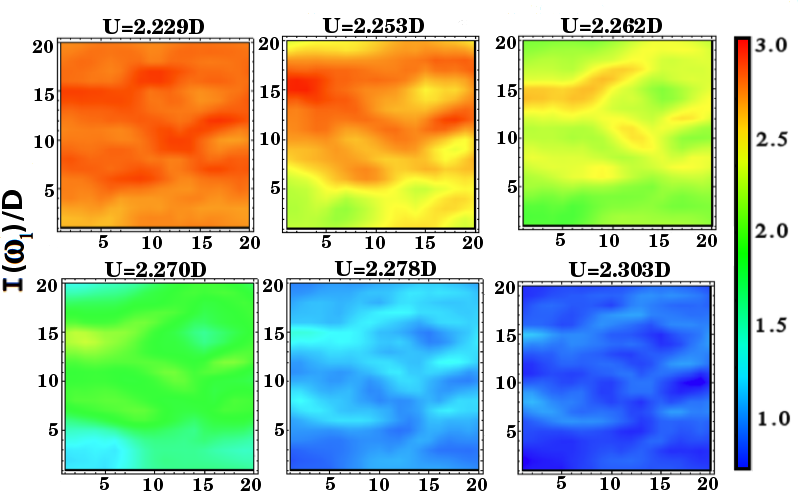}
\par\end{centering}
\centering{}\caption{\label{fig:current}Spatial mapping of the current in the vicinity
of the Mott transition for $T=0.024D$.}
\end{figure}

Using Eq (\ref{eq:R}) to find the values of the equivalent resistors
in the square lattice and Eq. (\ref{eq:ohm}) to calculate $I_{ij}$
between the nodes, it is possible to find the mean value of the current
in each node $I(i)=\frac{\Sigma_{j}|I_{ij}|}{N_{s}}$ , where $N_{s}$
is the number of resistors that are connected to a given node. In
Fig. \ref{fig:current} we present the results of the spatial mapping
of the current at $T=0.024D$ and $W=0.52D$, in the vicinity of the
Mott transition (\textcolor{black}{The Supplementary Material \citep{SMaterial}
shows solutions for other temperatures}). The red color represents
regions in which the current presents higher values. Low values of
the current are represented by the blue color and are associated with
insulating behavior. We note that the current is not uniform in the
system and has spatial fluctuations. However, its variations are mild
and the values \LyXZeroWidthSpace \LyXZeroWidthSpace decrease with
increasing $U$.

After calculating the value of the average current for each of lattice
site, we can find the average value over the complete network for
each interaction value $U$ in the vicinity of Mott transition. The
variation of the mean current is large for the case of $T=0.02D$
\textcolor{black}{(see Supplementary Material \citep{SMaterial},
for more details}). As the temperature increases the current in the
transition region still presents a noticeable change, but it is much
milder than in the case of $T=0.02D$. Since the external potentials
used in the calculations are fixed, the average current is a measure
of the conductance $G=\frac{I}{(\Phi-\Phi_{0})}$ . We note that,
although the conductance has a strong dependence on temperature in
the metallic regime, it is almost independent of $T$ in the insulating
regime. This is a consequence of the fact that $\mathrm{Im}G_{i}\left(i\omega_{1}\right)$
 cannot capture the exponential dependence with the temperature that
comes from the presence of the Mott gap.

\begin{figure}[H]
\begin{centering}
\includegraphics[scale=0.3]{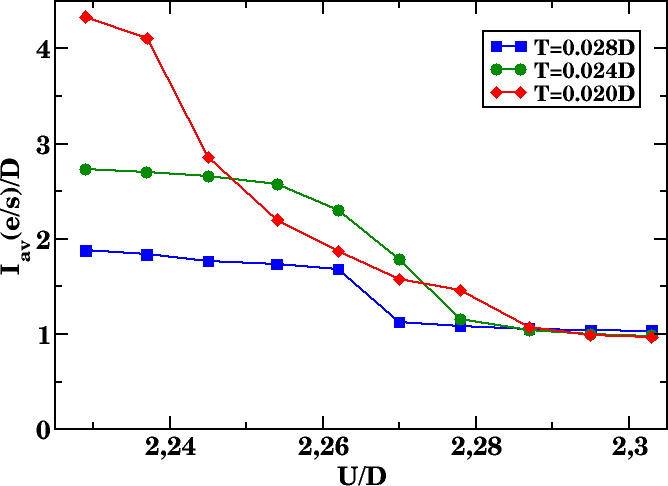}
\par\end{centering}
\centering{}\caption{\label{fig:average_current}Average current as a function of Coulomb
potential in the transition region of Mott for different temperature
values.}
\end{figure}

\section{\textit{\emph{Conclusions}}}

Using DMFT, in the clean case, and statDMFT, in the disordered case,
it was possible to analyze the Mott transition in Hubbard model in
a square lattice. In finite-sized lattices, disorder does not destroy
the first-order character of the transition with the accompanying
hysteresis loop and coexisting metallic and insulating solutions.
Since the local Green's function now has spatial fluctuations, however,
there is a different hysteresis loop on each site. The bundle of loops
shows an overall shift towards higher values of interactions when
compared with the clean case. The spatial pattern found shows clearly
the coexistence \emph{in each solution} of metallic as well as insulating
``bubbles''. As the system size increases, these different ``bubbles''
proliferate and point to a complete smearing of the first-order transition
in the thermodynamic limit, in complete agreement with the Imry-Ma
theorem. The statistics of local bare disorder fluctuations correlate
also reasonably well with metallic or insulating nature of the inhomogeneities,
which strengthens the link between the metal-insulator transition
in the disordered Hubbard model and the one of the random field Ising
model. Such a link had been previously emphasized in a scaling analysis
of the experimental results on the metal-insulator transition in VO$_{2}$.\citep{RFIMVO2}
Finally, we performed the first calculation of transport properties
within the statDMFT. This was possible only because the analyzed system
is a highly incoherent one, where $k_{F}l_{in}\sim a$, and the calculation
could be done through a mapping of the system onto a random network
of classical resistors. After this mapping, the global resistance
could be calculated and the temperature dependence in the metal is
in agreement with expectations. The same description fails in the
insulating case, however, where we expect to see an activated temperature
dependence.

\textcolor{black}{Our work offers a powerful theoretical perspective
on spatial inhomogeneities of disordered strongly correlated systems.}
In this sense, it is a welcome contribution to the description of
the detailed experimental results coming from recent nano-imaging
techniques. Besides the interplay of disorder and Mott physics explored
in this work, we envisage important directions for future work. In
the particularly well-studied example of VO$_{2}$\citep{triplepointVO2,LiuPRL}
as well as in other compounds\citep{strain} the coupling between
electronic and structural degrees of freedom is probably important
and could be incorporated. The inhomogeneous nanoscale patterns of
systems with competing orders, such as high-T$_{c}$ cuprates\citep{hightc}
and iron-based superconductors\citep{ironSC} would also benefit from
the kinds of insights gained from our approach. As these experimental
techniques mature, we expect more examples will be found where our
approach may prove useful.

\section{acknowledgments}

We acknowledge support by CNPq (Brazil) through Grants No. 307041/2017-4
and No. 590093/2011-8, Capes (Brazil) through grant 0899/2018 (E.M.),
NSF (USA) through Grant DMR-1822258 (V.D. and T-H.L), Texas Center
for Superconductivity at the University of Houston, \textcolor{black}{University
of Houston Health Research Institute,} the Center for Bioenergetics
at Houston Methodist Research Institute, and Leonardo Machado for
the helpful feedback (M.Y.S.V and J.H.M).

\bibliographystyle{apsrev}
\bibliography{all}

\end{document}